\input{aipcheck}

\documentclass[
    ,final            
  ]
  {aipproc}

\layoutstyle{6x9}

\begin{document}

\title{The Fate of Planets}


\author{Eva Villaver}{
  address={Universidad Aut\'onoma de Madrid, Dpto. F\'{\i}sica Te\'orica,
    M\'odulo 15, Facultad de Ciencias,  
Campus de Cantoblanco, 28049 Madrid, Spain. e-mail: eva.villaver@uam.es }
}

\begin{abstract}
As a star evolves off the Main Sequence, it endures major structural
changes that are capable of  
determining the fate of the planets orbiting it.  Throughout its evolution
along the Red Giant Branch,  
the star  
increases its radius by two orders of magnitude. Later, during the Asymptotic
Giant Branch, it loses most of  
its initial mass. Finally, during the Planetary Nebulae phase, it emits
intense radiation before ultimately  
beginning its fade as a white dwarf. We show how the several competing
processes (stellar mass-loss,  
gravitational and frictional drag, tidal forces, planet accretion and
evaporation) affect the survival of  
planets   
around evolved stars. 

\end{abstract}

\maketitle


\section{Planet's fate beyond the Main Sequence}

\subsection{The Red Giant Branch Phase}

Once the nuclear burning has been exhausted in the core, low- and
intermediate mass stars evolve into
the Red Giant Branch (RGB) phase. During the
RGB, hydrogen burning continues in a shell outside the helium core, which now,
devoid of energy sources, is contracting and heating up. The stellar
evolution timescales (in the absence of 
significant mass loss) are set by the rate of consumption of the nuclear
fuel. As the core
contracts the envelope expands and cools. The stellar effective temperature
decreases while the star's radius and luminosity increase.  

There are several competing processes that affect the orbital distance
between the star and the planet as the star evolves off the Main Sequence (MS): the changes
in the mass of both the planet and the star ($\dot{M_p}$ and $\dot{M_*}$
respectively), the gravitational and frictional 
drag ($F_g$ and $F_f$), and the tidal force.  To determine
the rate of change in the planet's orbit $a$ we consider a
planet of mass $M_p$ 
and radius $R_p$ moving with a velocity, $v$,  in a  circular orbit ($e = 0$)
around a star of mass $M_*$. The conservation of angular momentum  gives the
equation for the rate of 
change in the orbital radius of the planet (see, e.g., \citealt{Ale76,Ls84, Vl09}),  
\begin{equation} 
\left(\frac{\dot{a}}{a}\right) =-\frac{\dot{M_*}+\dot{M_p}}{M_*+M_p}-\frac{2}{M_p v}
\left[F_f+F_g\right]-\left(\frac{\dot{a}}{a}\right)_{t}~~,
\label{todo}
\end{equation}
where $({\dot{a}}/{a})_{t}$ is the rate of orbital decay due to the
tidal interaction.

It has been shown that for giant stars, which have massive convective
envelopes, the most efficient 
mechanism to produce tidal friction is turbulent viscosity 
\citep[e.g.,][]{Zah66,Zah77,Zah89}. The dissipation timescale is determined
by the effective eddy viscosity, with eddy velocities and length scales given
approximately by standard mixing length theory if convection transports most
of the energy flux \citep{Zah89,Vp95,Retal96}. The tidal term is given by  
\begin{equation}
\left(\frac{\dot{a}}{a}\right)_{t} = \frac{f}{\tau_d} \frac{M_\mathrm{env}}{M_*} q (1+q)
\left(\frac{R_*}{a}\right)^8~~,
\label{tidal}
\end{equation}
with $M_\mathrm{env}$ being the mass in the convective envelope, $q =
M_p/M_*$, and $\tau_d$ the eddy turnover timescale, given in the case of a
convective envelope \citep{Retal96},   
\begin{equation} 
 \tau_d = \left[\frac{M_\mathrm{env} (R_*-R_\mathrm{env})^2 }{3L_*}\right]^{1/3}~~,
\label{eddy}
\end{equation}
where $R_\mathrm{env}$ is the radius at the base of the convective
envelope. The term $f$ in Eq.~(\ref{tidal}) is a numerical factor obtained
from integrating the viscous dissipation of the tidal energy across the
convective zone. \cite{Zah89} used $f=1.01(\alpha/2)$ where $\alpha$ is the
mixing length parameter. \cite{Vp95} confirmed that observations are
consistent with $f\approx$1 as long as $\tau_d \ll P$ with $P$ being the orbital period. We
therefore used $f = (P / 2\tau_d)^2$ to account only for the convective cells
that can contribute to viscosity when  $\tau_d > P/2$, otherwise we take
$f=1$. It is important to note here that the initial value of the
eccentricity has little effect on the orbital decay rate \citep[e.g.,][]{Jac08}. 

In Fig.~1
we show the orbital evolution for 4 different initial orbital radius and for
planets orbiting stars with MS masses of 2 and 3 M$_{\rm \odot}$, solar metallicity, and a
mass-loss prescription with a Reimers parameter of $\eta_R= 0.6$
\citep{Rei}. The figure demonstrates the three possible
outcomes of orbital evolution for a Jupiter  
mass planet along the RGB \citep{Vl09}: (i) Beyond a certain initial orbital separation,
the orbital separation simply increases,  
due to systemic mass loss. (ii) There is a range of initial orbital
separations in which the orbit decays, but the  
planet avoids being engulfed. (iii) Inward from some critical, initial
orbital separation, the planet is engulfed  
mostly due to tidal interaction. Note that tidal interaction along the RGB
phase can determine a planet´s fate even  
at large distances beyond the stellar radius.

\begin{figure}
  \includegraphics[height=.35\textheight]{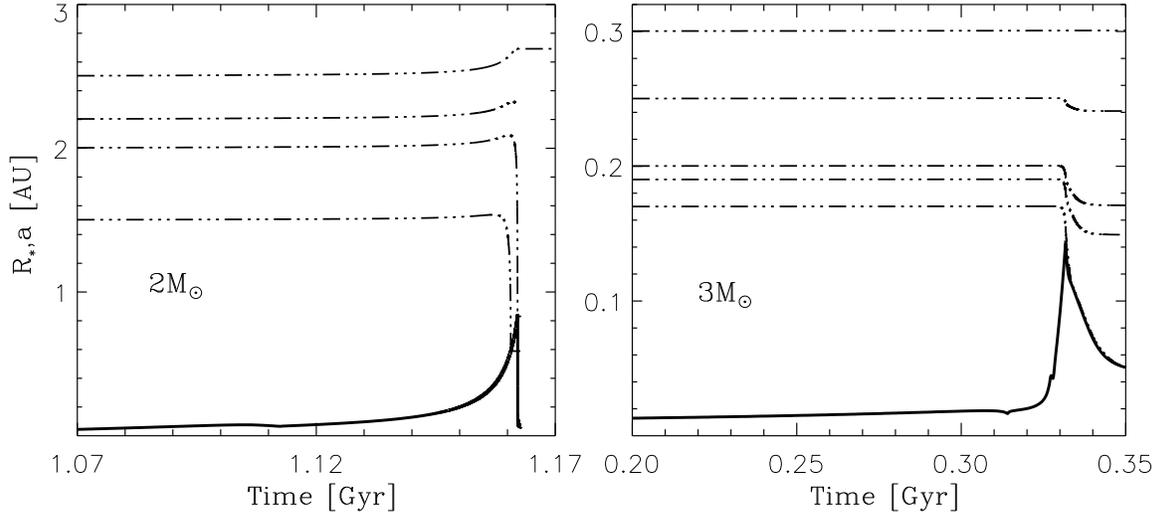}
  \caption{Evolution of the orbital separation of a planet with Jupiter's mass
 (dash-dotted line) compared with the evolution of the stellar radius (solid line) 
along the Red Giant Branch phase. The panel on the left is for a star with a main sequence 
mass of 2 M$_{\rm \odot}$ and the panel on the right for a  3 M$_{\rm \odot}$ star. The 
evolution of the orbital distance is shown for different initial orbits.}
\end{figure}

\begin{table}
\begin{tabular}{ccccc}
\hline
 \tablehead{1}{r}{b}{$M_*$}
 &\tablehead{1}{c}{b}{$R_*^\mathrm{max}$ [AU]} 
 &\tablehead{3}{c}{b}{ a$_\mathrm{min}$ [AU]}\\
&
&
\tablehead{1}{c}{b}{$M_p = M_{J}$} &
\tablehead{1}{c}{b}{$M_p = 3~M_{J}$}&
\tablehead{1}{c}{b}{$M_p = 5~M_{J}$}\\
\hline
1 M$_{\rm \odot}$ &1.10 &3.00 &3.40 &3.70\\
2 M$_{\rm \odot}$ &0.84 &2.10 &2.40 &2.50\\
3 M$_{\rm \odot}$ &0.14 &0.18 &0.23 &0.25\\
5 M$_{\rm \odot}$ &0.31 &0.45 &0.55 &0.60\\
\hline
\end{tabular}
\caption{Minimum Orbital Radius to Avoid Tidal Capture}
\label{tableorbit}
\end{table}

It is important to mention that the stellar structure enters into the
calculation of the tidal term. For the calculations presented here we have used
stellar models provided to us by Lionel Siess. These were calculated  based
on the stellar evolution code STAREVOL described in \cite{Siess06}.

\begin{figure}
  \includegraphics[height=.5\textheight]{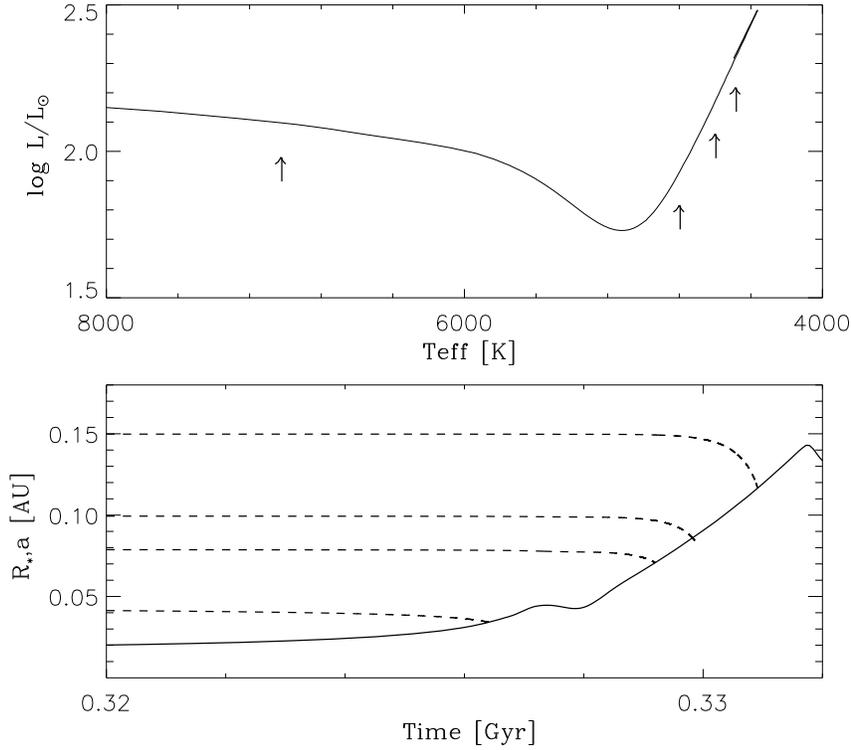}
  \caption{The top panel shows the evolution of a 3 M$_{\rm \odot}$ star
    along the Red Giant Branch 
phase on the HR diagram. The bottom panel shows the same as the right
panel of Fig.~1 but for  
5 M$_{\rm J}$ planets with different initial orbital separations. The
arrows show the location  
on the HR diagram at which the planet enters the stellar envelope. The
evolution of the stellar  
radius is shown as a solid line on the bottom panel.}
\end{figure}

In Fig.~2, the top panel
shows the evolution along the RGB of a 3 M$_{\rm \odot}$ star on the HR
diagram. The bottom panel 
shows the orbital evolution for a planet with mass 5~$M_J$ (dash-dotted line)
together with the evolution of the stellar 
radius (solid line). We selected small initial orbits to identify
at which points during the RGB these planets are swallowed by their stars
(marked by the location of the arrows).

Table~\ref{tableorbit} summarize some of our results, where we
list the minimum initial orbital distance for which a planet with a given
mass avoids being engulfed by the star.  The second column gives the maximum
radius reached by the star on the RGB, $R_*^\mathrm{max}$ (in AU), and the
following columns give the minimum orbital distances (in AU) at which planets
with masses of 1, 3, and 5~M$_{J}$ (respectively)  avoid being engulfed.

For
all the initial orbits that satisfy the condition $a_o \leq 
R_*^\mathrm{max}$ the planet gets engulfed by the star at same point before
the end of the RGB phase. The more massive the planet the stronger is the
tidal interaction with the star, and therefore the sooner the orbit decays to
meet the stellar radius.  

We would like to highlight that the tidal ``capture'' radius increases with
the planet's mass; a Jupiter-mass planet is captured by a 2~ M$_{\rm \odot}$ star
if it starts at an initial orbit of 2.1~AU while a 5~$M_J$ planet will be
engulfed by the star if it has an initial orbit $a_o < 2.5$~AU. Moreover, the
tidal capture radius decreases with increasing stellar mass; for a 5~$M_J$
planet the initial orbit has to be larger than $a_o \approx 3 \times
R_*^\mathrm{max}$ to avoid tidal capture around a 2~ M$_{\rm \odot}$ star, while it
has to be larger than $a_o \approx 2 \times R_*^\mathrm{max}$ to avoid tidal
capture around a M$_{\rm \odot}$  star. 

At large distances from the star, the densities involved are low, and the
drag terms associated with the forces F$_f$ and F$_g$ in Eq.~(\ref{todo})
play a negligible role in the evolution of the orbit. Moreover, since the
accretion rate onto the planet is always small compared to the stellar
mass-loss rate, the first term in Eq.~(\ref{todo}) is dominated by
$\dot{M_*}$. The temporal behavior of the orbit is then mostly governed by
the relative 
importance of the terms associated with the stellar mass-loss $\dot{M_*}$ and
the tidal interaction $(\dot{a}/a)_t$. Note also that the peak RGB mass-loss rates are
higher for lower-mass 
stars, and the lowest mass stars also reach the largest radius at the tip of the RGB.    
 
Red giant mass-loss rates are somewhat uncertain. It is important to
emphasize that (for the work presented here)  we have estimated the mass-loss
rate using the Reimers prescription 
with $\eta_R= 0.6$. This seems to reproduce fairly well the observations of
individual RGB stars. 
Uncertainties in both stellar evolution and mass-loss will directly influence
the planet's survival, a different set of evolutionary models using different
mass-loss rates will change the planet's fate. An increase in stellar
mass-loss will move the orbit outwards to the point where tidal forces are no
longer effective. Higher mass-loss will also have the effect of decreasing the
evolutionary timescales during this phase and might influence as well the
maximum stellar radius.

\subsection{The Asymptotic Giant Branch Phase}
The major
structural changes 
in the post-main sequence evolution of low- and  intermediate-mass stars occur
during the RGB and Asymptotic Giant Branch (AGB) phases. During the RGB and
AGB the stellar effective 
temperature 
is always lower than its main sequence value and therefore it has no influence
on the planet's survival. However, it is during the late AGB evolution, the
so-called  thermal-pulsing AGB phase, that a planet's orbit will be most
influenced, since during this phase the star loses most of its initial mass and
reaches its maximum radius.

If a planet becomes engulfed by the stellar envelope (it could be along the
RGB or the AGB phase) it can spiral-in and evaporate totally; or it can
accrete mass and become a close low-mass companion to the star. It is
important to mention that even 
if a planet gets inside the stellar envelope it may survive under certain
conditions. In \cite{Vl07} we give an estimate of
the maximum planet mass that can be evaporated inside an AGB envelope,  0.014 M$_{\rm \odot}$
~or 15~$M_J$. This 
estimate is very uncertain given it was
obtained by  equating the location of the evaporation region (where the local
sound speed in the stellar envelope matches the escape velocity from the
planet's surface) to the energy required to expel the envelope
\citep{Sok:96,Sok:98,Nt:98}.  The value of this maximum mass is very debatable
because it depends on several factors, such as the efficiency of envelope
ejection \citep{Pzy:98}, which are
largely unknown.  

The structural changes that an AGB star undergoes in response to the
dissipation of a planet in its interior are complex and have been extensively
explored by \cite{SL99a,SL99b}. So far the details of the destruction
of such a planet within the stellar envelope of an AGB star have not been
studied in detail. 

The final orbit reached by the planet at the end of the AGB phase is simply
related to the stellar initial-to final mass relation for planets in orbits that {\it
  avoid}  
engulfment along the AGB phase. The initial mass of
the star, the higher the amount of mass lost during the AGB
phase and hence the larger the planet's orbital expansion. Planets
reach a final orbital distance at the end of the AGB phase,
determined by multiplying the initial orbit by  $M_*/M_{WD}$ where $M_{WD}$
is the white dwarf mass. The orbital expansion
factors for white dwarf progenitor stars are given in \cite{Vl07}. 

Another effect to consider is whether the planet
becomes  unbound due to the change in mass of the central star.  Unbinding can
be expected if the stellar mass-loss timescale, $\tau_{mass-loss}$, satisfies
$\tau_{mass-loss} < \tau_{dyn}$, where $\tau_{mass-loss}$ is given by
\begin{equation}
\tau_{mass-loss} \sim \frac{M_*}{\dot{M}_*}
\end{equation} 
and the dynamical timescale,$\tau_{dyn}$, by
\begin{equation}
\tau_{dyn} \sim [\frac{r^3}{G(M_*+M_{p})}]^{1/2},
\end{equation}
with $\dot{M}_*$ being the stellar mass-loss rate, $M_p$ the planet's mass,
and G the gravitational constant.  
Given that $\tau_{dyn}\sim$50\,yr while the shortest mass-loss timescale is
$\tau_{mass-loss} \sim 10^5$\,yr, it is very unlikely that a planet will
become unbound due to the decrease in the stellar mass during the AGB phase. 

Therefore, the orbit of the planet will expand due to  the heavy stellar
mass-loss rates experienced during the AGB evolution if the planet has
avoided being engulfed during both the RGB and AGB phases. Larger differences
between the initial and final mass of the star are experienced for the more
massive progenitors, causing the orbits of planets orbiting the more massive
stars (note that we are always referring to stars in the 1--5  M$_{\rm \odot}$~mass
range for which complete stellar evolution models exist) to be modified by
the larger factors (up to 5.5 times larger than the initial orbit). 

\subsection{The Planetary Nebulae Phase}

During the post-main sequence evolution of the star the stellar effective
temperature 
always remains lower than its main sequence value. However, once the star
leaves the AGB phase, the high mass-loss rate ceases and the remnant core
moves in the HR diagram  at constant luminosity toward higher effective
temperatures into the Planetary Nebulae (PN) stage, before the star reaches
the white dwarf (WD) cooling 
track. The planet's orbit is not expected to change further at this stage.
However, the main processes responsible for shaping PNe (high velocity winds)
and powering PNe line emission (high stellar effective temperatures) need to be
considered to establish the survival of a planet during this phase.

\begin{figure}
  \includegraphics[height=.5\textheight]{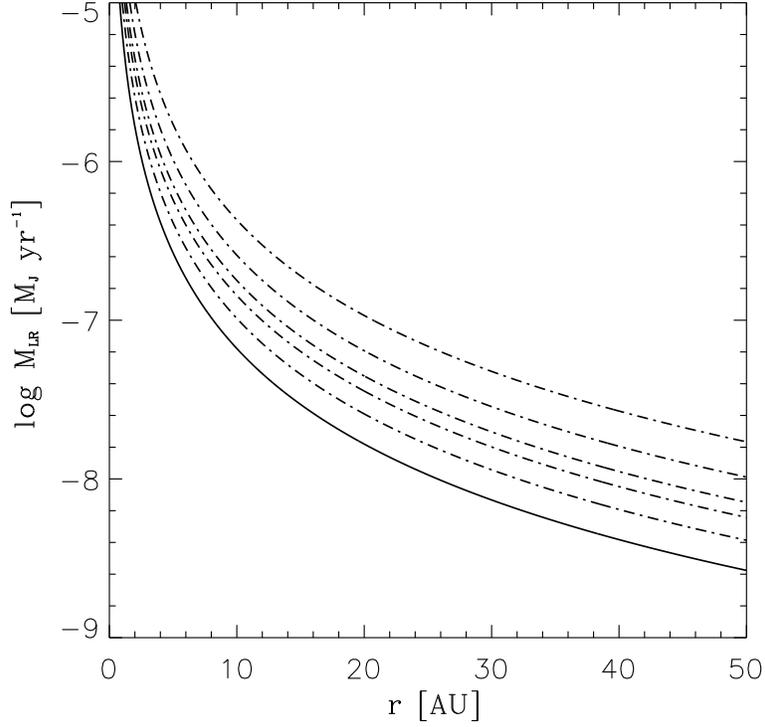}
  \caption{Mass-loss in logarithmic scale and M$_{\rm J} {\rm yr}^{-1}$
vs. orbital distance of a Jupiter-like planet under hydrodynamic scape
conditions. The different  
lines are for stellar MS masses 1, 1.5, 2, 2.5, 3.5 and 5 M$_{\rm \odot}$. The  
radiation field for the star has been taken at 30,000 {\rm yr} after the star
enters the Planetary Nebulae phase.} 
\end{figure}

The luminosity, mass, and timescale of evolution of the star during the PN
phase depend mostly on the stellar core mass. The stellar luminosity during this phase is
within the range 3.5 to 23 $\times 10^3$ L$_{\rm \odot}$~(for the lowest 0.56 and highest
0.9 M$_{\rm \odot}$~mass remnant, respectively) and the stellar temperature can reach
100\,000--380\,000\,K (for the same core masses, respectively). The hydrogen
ionizing photon flux, which is of the order of 10$^{48}$~s$^{-1}$
\citep{Vmg:02,Vgm02}, is 
responsible for the PNe  
ionized line emission. PNe central stars also emit very high velocity winds
(with speeds of a few thousands of \,km\,s$^{-1}$) which are driven by  the transfer of
photon momentum to the gas, through absorption by strong resonance lines. The
PN is largely shaped by the  interaction of this high 
velocity wind with the slowly ejected material during the AGB phase.  The survival of a gas
planet as the star evolves into the  PN phase strongly depends on the planet's
surface temperature, 
which ultimately determines whether or not high evaporation rates are set at
the planet's surface.  

\subsubsection{Planet's Evaporation Rates}

As a consequence of the high temperatures reached at the planet's upper atmosphere
it is expected that an outflow will develop due to the
absorption of the XUV radiation. High temperatures can cause the
outer layers of the planet to escape rapidly. Note that at the orbital
distances considered, the planet will be well within the typical inner
radius of the nebular shell (0.01 to 0.1 \,pc) \citep{Vmg:02} and therefore 
a decrease in the photon flux arriving at the planet's surface due to
absorption by the nebula is not expected. 

The problem of a general outflow from a stellar (or planetary) body can be 
described with the same set of equations used by \cite{Par:63}
to describe the solar wind. Although a complete treatment of the evaporative
wind 
requires the integration of the energy, mass, and momentum transfer equations,
we can estimate the outflowing particle flux $\Phi_{H}$  (e.g.\ \citealt{Wat:81}) by
equating the energy input $(\epsilon L_{\rm XUV}/4) \times (R_1/a)^2 $ to
the energy required for hydrogen to escape $GM_p m_H/R_p$, giving 
\begin{equation}
\Phi_{H}\simeq\frac{\epsilon L_{\rm XUV} R_1^2 R_p}{4 a^2 G M_p m_H}
\label{loss}
\end{equation}
\noindent
where $R_1$ is the planet's radius where most of the XUV radiation is absorbed,
defined as the level where the optical depth is unity, $a$ is the orbital
distance, $R_p$ and $M_p$ are the planet's radius and mass respectively and
$ \epsilon L_{\rm XUV}$ the fraction of the stellar XUV luminosity that is
converted into thermal energy. The most uncertain parameter in Eq.~\ref{loss}
is $R_1$, or ~$\xi = R_1/R_p$, 
since its determination requires the full solution of the hydrodynamical escape problem with
radiation transfer in a strongly externally heated atmosphere.
Generally, simulations produce higher escape rates than the ones 
given by Eq.~\ref{loss} for $\xi$ = ~3 as the density increases. This is the
result of the higher total amount of energy  absorbed in an extended
atmosphere, as opposed to that absorbed in a single layer (the approximation
used to obtain Eq.~\ref{loss}; see \citealt{Vl07}).

\begin{figure}
  \includegraphics[height=.5\textheight]{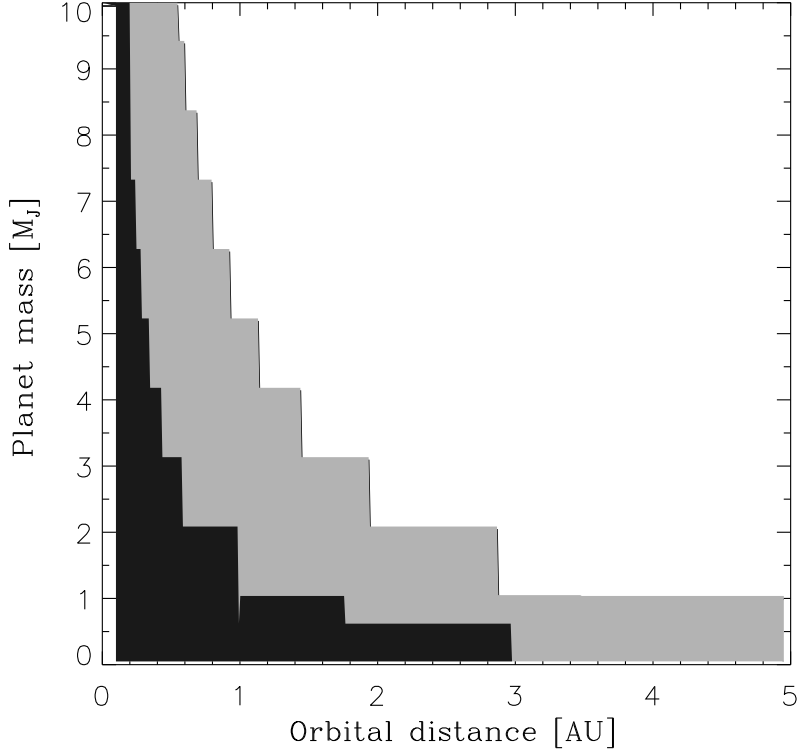}
  \caption{The dark-gray area represent the region where a gas planet loses
    50\% of its mass when orbiting a 0.56  
M$_{\rm \odot}$. The same is for the light-gray area but using a different
parameter for the calculation of evaporation rates.} 
\end{figure}

The mass-loss rates (obtained by using $\xi~=$~3 in  Eq.~\ref{loss})
from a 1~M$_J$~planet for central stars (at  36\,000 \,K) with different masses
are plotted in Fig.~3, versus the orbital distance. 
The different
lines account for the different central star masses considered, with the
mass-loss 
rates increasing with the stellar mass. 

It is important to note that none of the calculations in the literature for
exoplanets orbiting MS solar-mass stars includes heating
rates as 
high as the ones expected for a planet orbiting a PN central star, and as it
has been shown the inclusion of X-ray irradiation from the
star strongly increases the heating in planetary exospheres.
The mass-loss rates given in Fig.~3 should be considered
merely as order of magnitude approximations to the
actual mass-loss rates from a planet exposed to a PN central star. In addition,
it is very likely that the planet will inflate as radiation is transformed into
heat inside its atmosphere, which will lead to further increase in the
planet's evaporation rate with our approach.  An appropriate determination 
of the escape rate will require a solution of the hydrodynamic escape
equations for the case under consideration.

In Fig.~4 we show the region (on the planet mass versus orbital
distance plane) inside which Jupiter-like planets will be destroyed, 
as the star evolves into the white dwarf phase  orbiting a 0.56 M$_{\rm
  \odot}$~star (which 
correspond to 1 M$_{\rm \odot}$ main sequence mass). The dark gray and gray
shaded areas (computed for $\xi$=3 and 10 respectively, see \citealt{Vl07})
represent the regions 
for which planets will lose 50 \% of their mass before the star enters the
white dwarf cooling track due to the evaporation caused by thermal heating. 
Note that the region for which the planet will undergo total evaporation due
to thermal heating is inside the AGB stellar radius for the 0.9 M$_{\rm
  \odot}$ white dwarf.
\section{Planets Observed Around Evolved stars}

\subsubsection{ Giant stars}
To date, the frequency of planets around massive stars can only be determined
by searching stars that have left the MS. In only a few years Doppler
techniques applied to sub-giants and clump giants has succeeded in
discovering more than 40 planets around stars with masses between 1.2 and 4 M$_{\rm
  \odot}$ \citep{Doli07,Doli09,Frin02,Hat03,Hat05,Hat06,Joh07a,Joh07b,Joh08,Liu07,Lm07,
 Nied07,Nied09,Ref06,Sat07,Sat08,Set03,Set05}. 

Despite the small number statistics these surveys have revealed important clues 
about the planet distribution among the more massive evolved stars. First, it 
seems that there is a deficiency of  of close-in planets orbiting evolved stars 
with masses $M > 1.3 M_{\rm \odot}$ \citep{Joh07b,Sat08,Wri09} 
despite the fact that these planets are found around
$\approx$20 \% of the MS stars with $\approx$ 1 M$_{\rm
  \odot}$. Second, the
frequency of planets seems also to be higher around intermediate-mass stars 
\citep{Lm07,Joh07b}.   

In \cite{Vl09} we show that stellar evolution could explain
the observed distribution of the semi-major axes of planetary orbits  around
evolved stars (i.e., semi-major axis $>$\,0.5~AU). Despite the uncertainties
in stellar (and tidal) evolution theory when the
details of the orbital evolution are accurately calculated, tidal
interactions constitute a quite powerful mechanism, capable of capturing  
close-in planets into the envelope of evolved stars.

\subsubsection{White dwarfs}
Low-mass brown dwarfs and extrasolar planets in wide
orbits around white dwarfs would have temperatures lower than 500 K and
therefore could be detected via direct imaging in the Infrared
\citep{Bur}. Infrared searches for planets around white dwarfs
(e.g.\ \citealt{Mul,Go,Fa, Ho}) are
underway.  From these surveys it has has been found that approximately 2 \% of
all white dwarfs with cooling ages $\le$ 0.5 Gyr show evidence of an infrared excess
\citep{Fa9}. 

Moreover, disks  formed by the tidal disruption of an
                                asteroid have been proposed to explain the
                                 high anomalous photospheric-metal
                                 abundances found in some white dwarfs
                                 \citep{Be,Ki,Ga}. Furthermore, there
                                 are more than a dozen white dwarfs known to
                                 host circumstellar debris disks
                                 (e.g.\ \citep{Fa9,Ga,Ki}). 

 Planets around
                                 white dwarfs with masses $M_{\rm WD}
                                 \ge$ 0.7M$_{\rm \odot}$ 
  ~(that formed from single stars with masses $M_{MS} \ge$ 2.5 M$_{\rm
  \odot}$), are generally expected to be found at orbital radii r $\ge$ 15
  \,AU due to the effects of mass-loss on the AGB phase \cite{Vl07}. The sensitivity of
  the current surveys at such larger distances from the star has to be further explored.

\subsubsection{ Planets and sdB stars}  

Among the planets found orbiting evolved
stars, there are several systems that beg further investigation: e.g.\ the
planet of mass 3.2 $M_J$ 
 with an orbital separation of 1.7 \,AU 
orbiting the extreme horizontal branch star V391 Pegasi \citep{Si}. The star V391 Pegasi
lost its hydrogen-rich envelope (for reasons that are not well understood) at
the end of the Red Giant Branch leaving behind a hot B-type sub-dwarf (sdB)
with a surface temperature of 30,000 K. 

More recently, \cite{Ge}
has found  a planet orbiting the sdB HD 149382 at a distance of only about
five solar radii. At that distance, the planet must have survived engulfment
in the red giant envelope. 

Serendipitous discoveries of two substellar
companions around the eclipsing sdB binary HW Vir at distances of 3.6 AU and
5.3 AU (Lee et al.\ 2009) and one brown dwarf around the similar system HS
0705+6700 with a separation of < 3.6 \,AU \citep{Qi} followed
recently. It seems more than plausible that the presence of a planetary
companion can trigger envelope ejection and enabled the sdB star to form. Hot
sub-dwarfs have been identified as the sources of the unexpected ultraviolet
(UV) emission in elliptical galaxies, but the formation of these stars is not
fully understood (see e.g.~\citealt{He}). 

\section{Summary}
The conditions for planet survival as the star evolves off the main sequence
depend on the initial mass of the star. Most of the MS
close-in planets will most likely be destroyed as they get engulfed by
the star during the RGB and
AGB phase. As the star leaves the AGB and enters the PN
phase, high effective temperatures at very high luminosities set up an
evaporation flux at the  surface of the planet. At certain orbital distances
the evaporation rates are high enough to cause a total destruction of the
planet. By integrating the evaporation rates as the star evolves during the PN
into the white dwarf cooling track, Jupiter-like planets will be
destroyed if they remain at orbital distances \,r$\le$ 5\,AU from a low mass
white dwarf (M$_{\rm WD} \le$ 0.63 M$_{\rm \odot}$) and large planet ablation is expected
up to 10\,AU. In particular, Jupiter in our own Solar system is barely
expected to survive. More massive stars evolve very fast during the  PN phase
and do 
not maintain high evaporation rates long enough to cause planet destruction,
unless the planets were to be found at small orbital distances (r$\le$2.5
\,AU). However, we have shown that planets orbiting the more massive PN central stars
cannot be found at small orbital distances: if a planet orbiting a 0.9  M$_{\rm \odot}$
progenitor is to survive AGB engulfment then its orbit has to be beyond r$\ge$
29 \,AU. 

We find that the evolution of the star alone can quantitatively explain the
observed lack of close-in planets around evolved stars even allowing for the
uncertainties associated with mechanisms such as mass loss along the RGB or
tidal-interaction theory. 
Along similar lines, since we find a high probability of tidal capture of the
planet by evolved stars, the higher frequency of planets observed around
intermediate-mass stars \citep{Lm07,Joh07b} seems to imply that the 
efficiency of planet formation must be considerably higher for more massive
stars, compared to their solar analogous.

\end{document}